\documentclass{article}
\usepackage{spconf,graphicx}
\usepackage{amsmath,amssymb,amsfonts}
\usepackage{mathrsfs}
\usepackage{color}
\usepackage{bm}
\usepackage{subfigure}
\usepackage{dblfloatfix}
\usepackage{arydshln}
\usepackage{enumitem}

\title{Dynamic Alignment Mask CTC: IMPROVED MASK CTC WITH ALIGNED CROSS ENTROPY}
\name{Xulong Zhang$^{1\dagger}$, Haobin Tang$^{1,2\dagger}$, Jianzong Wang$^{1\ast}$, Ning Cheng$^1$, Jian Luo$^1$, Jing Xiao$^1$\thanks{$\dagger$  Equal contribution.}\thanks{$^\ast$Corresponding author: Jianzong Wang, jzwang@188.com.}}
\address{$^1$Ping An Technology (Shenzhen) Co., Ltd.\\$^2$University of Science and Technology of China}

\begin{document}
%\ninept
%
\maketitle
\begin{abstract}
%In real-world deployment of speech recognition, fast inference speed is critical, especially in the online streaming scenarios.
Because of predicting all the target tokens in parallel, the non-autoregressive models greatly improve the decoding efficiency of speech recognition compared with traditional autoregressive models. In this work, we present dynamic alignment Mask CTC, introducing two methods: (1) Aligned Cross Entropy (AXE), finding the monotonic alignment that minimizes the cross-entropy loss through dynamic programming, (2) Dynamic Rectification, creating new training samples by replacing some masks with model predicted tokens. The AXE ignores the absolute position alignment between prediction and ground truth sentence and focuses on tokens matching in relative order. The dynamic rectification method makes the model capable of simulating the non-mask but possible wrong tokens, even if they have high confidence. Our experiments on WSJ dataset demonstrated that not only AXE loss but also the rectification method could improve the WER performance of Mask CTC.
\end{abstract}

\section{Introduction}
\label{sec:intro}

Recently, end-to-end automated speech recognition (ASR) systems have received a lot of interest. There have been some prior efforts to realize non-autoregressive (NAR) models~\cite{Song2021Non,Fan2021CASS,Tian2020Spike,Bai2020Listen,Majumdar2021Citrinet}. The most common method is Connectionist Temporal Classification (CTC)~\cite{graves2006connectionist}. Wenet\cite{zhang2021wenet} applies CTC beam search to generate $N$-best candidates and uses the decoder to rescore the probabilities of these candidates, generating the final token sentence. Because the sentence length of candidates is fixed in the rescored process, it makes Wenet have a non-autoregressive structure and fast decoding speed. Conditional masked language model (CMLM) is another method adopted in non-autoregressive ASR models. CMLM has proved to be an effective model in various natural language processing (NLP) tasks~\cite{Devlin2019BERT,ghazvininejad2019mask,ghazvininejad2020semi}, and has been introduced into ASR tasks recently. For instance, Imputer~\cite{chan2020imputer} optimizes the potential alignment of the CTC iteratively by predicting the frame-level mask of the input speech. Compared with Imputer, Mask CTC~\cite{higuchi2020mask} generates a shorter sequence by refining CTC output depending on mask token prediction. At the inference stage, the target sentence is initialized using the greedy CTC output. And then based on the CTC probability, tokens with low confidence are masked.
% In each iteration, the masked tokens are optimized and conditioned on the other observed tokens and the acoustic features. 
In each iteration, optimization conditioned on the other observed tokens and the acoustic features is performed on the masked tokens.
DLP Mask CTC~\cite{higuchi2020improved} is an improved version of Mask CTC, which uses an additional predictor to estimate the length of local masks. In addition, DLP Mask CTC also employed Conformer~\cite{Gulati2020Conformer} to enhance the encoder network architecture. DLP Mask CTC obtains a substantial recognition accuracy improvement over Mask CTC.

% \textbf{High Confidence Prediction Problem}
Despite achieving a high inference efficiency, the above non-autoregressive models still encounter some tough problems. One is that the decoder network of NAR model is usually trained on cross entropy (CE) loss, but the CE loss might be too strict for NAR model training. Cross entropy requires the force-alignment between the model predictions and ground truth tokens. It means that small shift in predicted tokens will result in large loss penalty, even if the content of tokens matches very well. Aligned cross entropy (AXE)~\cite{Ghazvininejad2020Aligned} was proposed as a relaxed loss function for training non-autoregressive natural language processing (NLP) models. %AXE defines three local update operators: (1) $\text{align}$, (2) $\text{skip\_prediction}$, and (3) $\text{skip\_target}$, to handle different mapping situations between ground truth tokens and model predictions. In addition, AXE uses a differentiable dynamic programing to calculate the optimal alignment efficiently.
In machine translation tasks, CMLM models trained with AXE loss could significantly outperform the models trained with traditional CE loss~\cite{Ghazvininejad2020Aligned}. In this work, we introduce the AXE loss to the decoder training of Mask CTC. We think AXE loss could make the model focus on the tokens matching instead of tokens ordering to improve the recognition accuracy of NAR speech recognition models.

Another issue is that the decoder input of Mask CTC is the greedy CTC search at the inference stage, while the ground truth sentence is inputted to the decoder at the training stage. This causes a mismatch between the training and inference. The high confidence tokens of CTC search are reserved as non-mask tokens. However, they may also have errors. These non-mask tokens could not be refined and influence the filling of neighbor mask tokens. In this paper, we propose a dynamic rectification method to alleviate this problem. 

\begin{table*}[!b]
	\centering
	\caption{Illustration of Mask Method and Dynamic Rectification}
	\label{tab1}
	\scalebox{0.95} {
		\begin{tabular}{p{1.5cm}|cccccccccc}
			\hline\hline
			\bm{$Y$} & This & is & dynamic & alignment & \textbf{mask} & CTC & non & autoregressive & speech & recognition \\
% 			\bm{$M_{mask}$} & - & $\left \langle mask \right \rangle$ & - & $\left \langle mask \right \rangle$ & $\left \langle mask \right \rangle$ & - & - & $\left \langle mask \right \rangle$ & $\left \langle mask \right \rangle$ & - \\
			\bm{$Y_{mask}$} & This & $\left \langle mask \right \rangle$ & dynamic & $\left \langle mask \right \rangle$ & $\left \langle mask \right \rangle$ & CTC & non & $\left \langle mask \right \rangle$ & $\left \langle mask \right \rangle$ & recognition \\
			\bm{$\widetilde{Y}$} & This & is & dynamic & alignment & \textbf{task} & CTC & non & autoregressive & speech & recognition \\
% 			\bm{$M_{rec}$} & - & - & $\left \langle mask \right \rangle$ & $\left \langle mask \right \rangle$ & - & - & $\left \langle mask \right \rangle$ & - & - & - \\
			\bm{$Y_{rec}$} & This & is & $\left \langle mask \right \rangle$ & $\left \langle mask \right \rangle$ & \textbf{task} & CTC & $\left \langle mask \right \rangle$ & autoregressive & speech & recognition \\
			%			\textbf{shrink} $\left \langle mask \right \rangle$ & This & $\left \langle mask \right \rangle$ & dynamic & $\left \langle mask \right \rangle$ & CTC & non & $\left \langle mask \right \rangle$ & recognition & - & - \\
			%			\textbf{add} $\left \langle eos \right \rangle$ & This & $\left \langle mask \right \rangle$ & dynamic & $\left \langle mask \right \rangle$ & CTC & non & $\left \langle mask \right \rangle$ & recognition & $\left \langle eos \right \rangle$  & $\left \langle eos \right \rangle$  \\
			\hline\hline
		\end{tabular}
	}
\end{table*}

\section{Proposed Method}
\label{sec:method}

\begin{figure}[ht]
	\begin{center}
	% \vspace{-2ex}
		\centerline{\includegraphics[width=0.9\columnwidth]{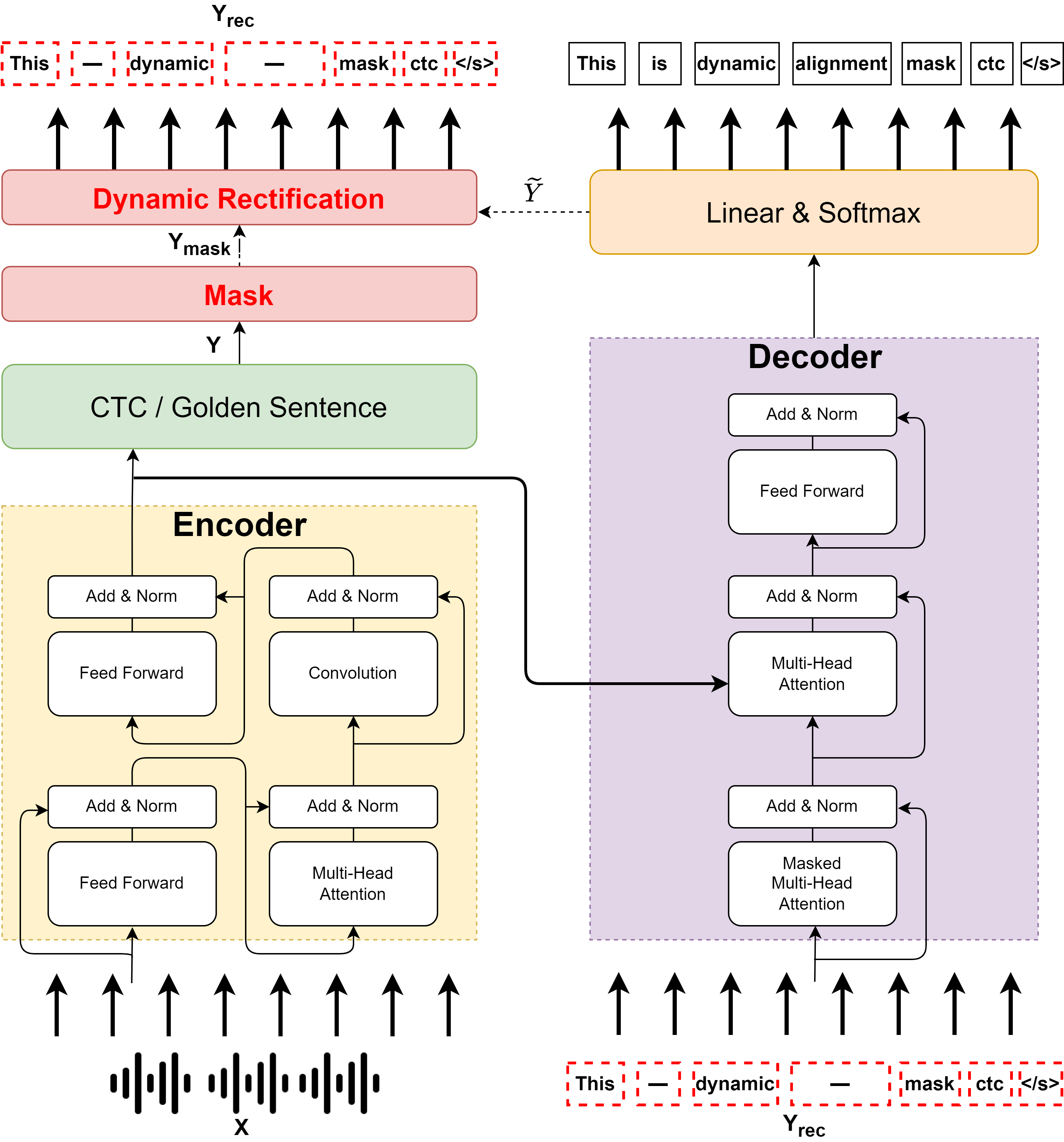}}
		\caption{Conformer Architecture of Dynamic Alignment Mask CTC with Mask and Dynamic Rectification Methods}
		\label{fig1}
	\end{center}
 \vspace{-2em}
\end{figure}
We propose a non-autoregressive end-to-end automatic speech recognition framework. Our works are based on Mask CTC~\cite{higuchi2020mask}. We try to improve Mask CTC by two methods: (1) aligned cross entropy $\theta_{axe}$, finding a monotonic alignment to minimize the cross entropy loss on masked tokens, (2) dynamic rectification $\theta_{rec}$, simulating the high confidence unmasked tokens by using model predictions, to produce new training samples. Aligned cross entropy is a relaxed loss, which ignores the absolute positions and focuses on the relative order and tokens matching. Dynamic rectification uses predicted tokens as the input of model decoder, closing to the greedy CTC search result at the inference procedure.

\subsection{Model Architecture}
\label{ssec:architecture}

As Figure~\ref{fig1} depicts, the proposed dynamic alignment Mask CTC is encoder-decoder model architecture built on Conformer~\cite{Gulati2020Conformer} (or Transformer~\cite{vaswani2017attention}) blocks. The encoder transforms a speech acoustic feature sequence $X = (x_1,x_2,...,x_T)$ to a hidden representation $H$. The decoder network outputs a text label sentence $Y = (y_1,y_2,...,y_S)$. We apply CTC to the encoder output $H$, and the Aligned Cross Entropy (AXE) objective is used to train the decoder.
% and the decoder is trained via the Aligned Cross Entropy (AXE) objective. 
% CTC predicts a frame-level alignment $A$ between the input sequence $X$ and the output sentence $Y$. 
Frame-level alignment $A$ is predicted by CTC between the input sequence $X$ and the output sentence $Y$. The definition of the CTC loss~\cite{graves2006connectionist} is stated below:
% The CTC loss~\cite{graves2006connectionist} is defined as follows:
\begin{equation}
\mathcal{L}_{ctc} \triangleq -\ln~P_{ctc}(Y|X) = -\ln~\sum_{A \in\beta^{-1}(Y)}P(A|X)
\end{equation}
% where $P(Y|X)$ is the probability of the corresponding text label at frame step $t$ of encoder output. 
where $P(Y|X)$ is the probability over all possible paths. 
$\beta^{-1}(Y)$
returns all possible alignments compatible with $Y$.  

At the training stage of dynamic alignment Mask CTC, we firstly applied mask method $\theta_{mask}$ on ground truth target sentence $Y$ to produce $Y_{mask}=\theta_{mask}(Y)$. Secondly, dynamic rectification method $\theta_{rec}$ modifies the decoder input sentence to $Y_{rec}=\theta_{rec}(Y_{mask})=\theta_{rec}(\theta_{mask}(Y))$. Then, the decoder predicts the whole label sentence $Y$ conditioning on the input audio $X$ and modified token sentence $Y_{rec}$ as follows:
\begin{equation}
\label{axe}
\begin{aligned}
\mathcal{L}_{axe} &\triangleq -\ln~P_{axe}(Y|Y_{rec},X) \\
&= -\sum_{i=1}^{S}\ln~P_{\alpha(i)}(y_i|Y_{rec},X)-\sum_{k}\ln~P_k(\epsilon|Y_{rec},X)
\end{aligned}
%\begin{aligned}
%\mathcal{L}_{cmlm} &\triangleq -\ln~P_{cmlm}(Y|Y_{rec},X) \\
%&= -\ln~\prod_{s}P(y_s|Y_{rec},X)
%\end{aligned}
\end{equation}
Here, we introduce the Aligned Cross Entropy (AXE) loss~\cite{Ghazvininejad2020Aligned} to the training of decoder instead of CMLM loss in Mask CTC. In which, $P_{\alpha(i)}(y_i|Y_{rec},X)$ denotes
 an aligned cross entropy between $Y$ and $\widetilde{Y}$. $P_k(\epsilon|Y_{rec},X)$ denotes a penalty for unaligned predictions, which is noted as a special token $\epsilon$. $k$ is the index of unaligned tokens in $\widetilde{Y}$.
Finally, through the CTC loss $\mathcal{L}_{ctc}$ and AXE loss $\mathcal{L}_{axe}$, the overall loss $\mathcal{L}$ of dynamic alignment Mask CTC is calculated with scale factor $\lambda \in [0,1]$.
% 0 \leq \lambda \leq 1$.
\begin{equation}
\mathcal{L} = \lambda~\mathcal{L}_{ctc} + (1-\lambda)~\mathcal{L}_{axe}
\end{equation}

\subsection{Mask Method and Dynamic Rectification}
\label{ssec:mask}
As illustrated in Table~\ref{tab1}, the mask method starts with an input text sentence $Y$. For training, $Y$ is ground truth sentence of labeled data. For inference, $Y$ is the greedy CTC search result without using beam search like Mask CTC. At training stage, the ground truth tokens are randomly substituted with $\left \langle mask \right \rangle$. The number of mask tokens is sampled from uniform distribution between $1$ to $L_{mask}$. During inference, we mask the positions with low confidence below the threshold $P_{thres}$. Therefore, the masked sentence $Y_{mask}$ could be obtained through this mask method $\theta_{mask}$ by $Y_{mask}=\theta_{mask}(Y)$.

% \subsection{Dynamic Rectification}
% \label{ssec:rectification}
During training, we use current best ASR model to predict text sentence $\widetilde{Y}$ from the mask method output $Y_{mask}$. $\widetilde{Y}$ will fill the masks of $Y_{mask}$ with the highest confidence in these mask locations. After that, the dynamic rectification method masks the text sentence $Y_{rec}$ again. The number of second rectification masks is between $1$ to $L_{rec}$. Through dynamic rectification, the output sentence $Y_{rec}$ will not only contains mask tokens but also may contain wrong predicted tokens with high confidence. It simulates the result of greedy CTC at the inference stage and alleviates the mismatch problem of decoder input between training and inference.

\subsection{Aligned Cross Entropy}
\label{ssec:axe}
The AXE loss is the minimum over all possible monotonic alignments $\alpha:\widetilde{Y}\rightarrow Y$ of the conditional cross entropy loss as shown in Eq.~\ref{axe}. Dynamic programming is used by AXE to determine the optimal alignment between the current prediction $\widetilde{Y}_j$ and ground truth token $Y_i$.
The matrix $M_{i,j}$ ($i,j\in[1:S]$) represents the minimal AXE loss value for this optimal alignment. 
% Because the length of decoder output is equal with the ground truth sentence length $S$ in dynamic alignment Mask CTC, we could obtain the overall AXE loss $\mathcal{L}_{axe}$ through the matrix value $A_{S,S}$.
% \begin{method}[ht]
% 	\caption{Aligned Cross Entropy $\theta_{axe}$}
% 	\label{method2}
% 	\begin{methodic}[1]
% 		\STATE {\bfseries Input:} ground truth tokens $Y$, predictions $\widetilde{Y}$
% 		\STATE $A_{0,0} = 0$
% 		\FOR{$i=1$ {\bfseries to} $S$}
% 		\STATE $A_{i,0} = A_{i-1,0} - \delta \cdot \log P(\widetilde{Y}_1|Y_i, X)$
% 		\ENDFOR
% 		\FOR{$j=1$ {\bfseries to} $S$}
% 		\STATE $A_{0,j} = A_{0,j-1} - \log P(\varepsilon|Y_i, X))$
% 		\ENDFOR
% 		\FOR{$i=1$ {\bfseries to} $S$}
% 		\FOR{$j=1$ {\bfseries to} $S$}
% 		\STATE $\text{align} = A_{i-1,j-1} - \log P(\widetilde{Y}_j|Y_i, X)$
% 		\STATE $\text{skip\_prediction} = A_{i,j-1} - \log P(\varepsilon|Y_i, X)$
% 		\STATE $\text{skip\_target} = A_{i-1,j} - \delta \cdot \log P(\widetilde{Y}_j|Y_i, X)$
% 		\STATE $A_{i,j} = \min \{ \text{align}, \text{skip\_prediction}, \text{skip\_target} \}$
% 		\ENDFOR
% 		\ENDFOR
% 		\STATE {\bfseries return} $A_{S,S}$
% 	\end{methodic}
% \end{method}
Three local update operators are defined in the dynamic programming of AXE loss~\cite{Ghazvininejad2020Aligned}: (1) $\text{align}$, aligning the current prediction $\widetilde{Y}_j$ and ground truth token $Y_i$ with probability $P(\widetilde{Y}_j|Y_i, X)$, (2) $\text{skip\_prediction}$, skipping the current prediction $\widetilde{Y}_j$ and inserting a special token $\varepsilon$ to the ground truth token $Y_i$, (3) $\text{skip\_target}$, skipping the current ground truth token $Y_i$ without incrementing the prediction $j$. This operation is penalized with the hyperparameter $\gamma$.
% $\delta$.
\begin{align}
&align = M_{i-1,j-1} - \log P(\widetilde{Y}_j|Y_i, X) \\
&skip\_prediction = M_{i,j-1} - \log P(\varepsilon|Y_i, X) \\
&skip\_target = M_{i-1,j} - \gamma \cdot \log P(\widetilde{Y}_j|Y_i, X)
\end{align}
$A_{i,j}$ is filled by taking the minimum from above three possible operators. 
% The cell $A_{S, S}$ will contain the cross entropy loss of the optimal alignment after dynamic programming. AXE loss relaxes the penalty for token order errors and focuses on the predicted token matching. 
After dynamic programming, the cross entropy loss of the optimal alignment will be present in the cell $M_{S, S}$. With AXE loss, the emphasis is placed on the expected token matching rather than the penalty for token order mistakes.
Otherwise, the traditional CE loss is a too strict criterion for small token shifts in $\left \langle mask \right \rangle$ locations.

\subsection{Decoding Strategy}
\label{ssec:decoding}
At inference stage of dynamic alignment Mask CTC, the ground truth sentence $Y$ is replaced by greedy search result of CTC decoding. We also applied mask method on $Y$ to obtain $Y_{mask}$. But we do not use dynamic rectification, which is only activated at training stage. Therefore, $Y_{mask}$ is directly as the input of decoder, to get the final predicted sentence $\widetilde{Y}$. We apply the iterative decoding methods with $K$ total decoding iterations. For each iteration, the decoder predicts the masked locations $\widetilde{y}_m$ in sentence $\widetilde{Y}$ as follows:
\begin{equation}
\widetilde{y}_m = \mathop{\arg\max}_{w}~P_{axe}(\widetilde{y}_m=w|Y_{mask},X)
\end{equation}
Then, top $C$ masked tokens with the highest probability are reserved, where $C$ is the average number of mask tokens for $K$ iterations. Other masked tokens are changed to masks for the next iteration. Until all the mask tokens are filled, the decoding method terminates.

%\subsection{Illustration}
%\label{ssec:illustration}
%
%In this section, we illustrated the mask and dynamic rectification method. As Table~\ref{tab1} depicts, we started with the ground truth sentence $Y$, and randomly mask some tokens to get $Y_{mask}$.
%%After shrinking consecutive $\left \langle mask \right \rangle$, $\left \langle eos \right \rangle$ are added at the end of $Y_{mask}$. The method forces the model aligned the shrinked masks and target tokens dynamicly.
%After that, $Y_{mask}$ is inputted into dynamic rectification method, as shown in Table~\ref{tab2}. We used current best model to predict $\widetilde{Y}$ based on $Y_{mask}$, and masked the mask tokens again. Therefore, the output sentence $Y_{rec}$ may has wrong predicted tokens (like ``task'' in $Y_{rec}$ instead of ``mask'' in ground truth sentence $Y$). Finally, the sentence $Y_{rec}$, ground truth sentence $Y$, and audio features $X$, compose the new training sample. The wrong predicted tokens of $Y_{rec}$ are aimed to simulate the high confident but possible wrong tokens of greedy CTC search result.

\section{Experiments}
\label{sec:exp}
%In this section, we evaluated our proposed methods in speech recognition task. We explored the choices of mask probability, threshold ratio, and decoding iterations. We also showed the convegence of model training, and visualized the dynamic programming of aligned cross entropy.

%\subsection{Dataset}
%\label{ssec:dataset}
%We performed all the experiments on the open vocabulary dataset WSJ, an English speech corpus of ASR task. The WSJ1~\cite{linguistic1994wsj1} and WSJ0~\cite{linguistic2007wsj0} are read speech on Wall Street Journal news text recorded by microphones. We used si284 for training, dev93 for validation, and eval92 for testing.
%\begin{itemize}
%	\setlength{\itemsep}{0pt}
%	\setlength{\parsep}{0pt}
%	\setlength{\parskip}{0pt}
%	\item \textbf{WSJ}: The WSJ1~\cite{linguistic1994wsj1} and WSJ0~\cite{linguistic2007wsj0} are read speech on Wall Street Journal news text recorded by microphones. We used si284 for training, dev93 for validation, and eval92 for testing.
%\end{itemize}

\subsection{Configuration}
\label{ssec:configuration}
We perform all the experiments on public speech corpus, WSJ1~\cite{linguistic1994wsj1} and WSJ0~\cite{linguistic2007wsj0}. 
% We use si284 for training, dev93 for validation, and eval92 for testing. 
We use si284, dev93, and eval92 for training, validation, and testing respectively.
All the experiments are conducted on the ESPNET2 toolkit~\cite{Watanabe2021ESPnet}.

For the network inputs, the audio data is encoded with $80$ mel-scale filterbank coefficients with $3$-dimensional pitch features extracted using Kaldi recipe. 
We use speed perturbation~\cite{Tom2015Audio} and SpecAugment~\cite{Daniel2019SpecAugment} as data augmentation techniques to avoid model overfitting. The Transformer encoder consists of $2$ CNN layers and $12$ self-attention layers. $12$ Conformer blocks make up the conformer encoder. The decoder consists of $6$ self-attention layers. The multi-head attention has $4$ heads, $256$ dimension, and $2048$ feedforward dimension. The dimension of decoder output is $65$ (including capital English letters, masks, and punctuations). The final model was derived by averaging the top $30$ models based on their validation accuracy after the training process.
% After training, the final model was obtained by averaging from the best $30$ models based on their validation accuracy. 
The scale factor $\lambda$ of CTC and AXE loss is set to $0.3$. During inference, the CTC confidence threshold $P_{thres}$ is $0.999$.

\subsection{Results}
\label{ssec:results}

We first explore the effectiveness of different modules of our dynamic alignment Mask CTC model as follows:
1) \textbf{Mask + CE}: The model trained with the joint CTC and CE loss. This is the baseline Mask CTC model;
2) \textbf{Mask + AXE}: The model which is only processed by mask method, and is trained on joint CTC and AXE loss;
3) \textbf{Mask + Rec + AXE}: We apply mask and dynamic rectification methods to create training samples and train this model on joint CTC and AXE loss.
% \begin{itemize}
% %	\setlength{\itemsep}{0pt}
% %	\setlength{\parsep}{0pt}
% %	\setlength{\parskip}{0pt}
% 	\item\textbf{Mask + CE}: The model trained with the joint CTC and CE loss. This is baseline Mask CTC model. %We choosed it as our baseline model.
% 	\item\textbf{Mask + AXE}: The model which is only processed by mask method, and is trained with joint CTC and AXE loss.
% 	\item\textbf{Mask + Rec + AXE}: We applied both mask and dynamic rectification methods to create training samples, and also trained this model on joint CTC and AXE loss.
% \end{itemize}

%All of the models are evaluated by WER (Word Error Rate) and RTF (Real Time Factor). As listed in Table~\ref{tab3}, we compared the results of our models with baseline Mask + CE model (we referenced the numbers of Mask CTC . When replacing the traditional CE loss with AXE loss, the WER result will decrease significantly. We also explored different choices of mask mask probability $\rho_{mask}$, and found that $\rho_{mask}=100\%$ still works best in our experiments.
%Our Mask + AXE model achieves relatively $4.8\%$ improvement on WER of dev93 validation set and $12.2\%$ on WER of eval92 testing set. This demonstrated that our models have better generalization ability with training on AXE loss. With dynamic rectification method, our Mask \& Rectification + AXE model could further improve the WER results on both dev93 and eval92 dataset. The mask probability of rectification $\rho_{rec}=100\%$ has the best performance at our experiments.
All models are evaluated by WER (Word Error Rate) and RTF (Real Time Factor). As listed in Table~\ref{tab3}, we compared the results of our models with baseline Mask + CE model. When replacing the traditional CE loss with AXE loss, the WER result of Mask + AXE model will decrease significantly. %We also explored different choices of mask mask probability $\rho_{mask}$, and found that $\rho_{mask}=100\%$ still works best in our experiments. Our Mask + AXE model achieves relatively $4.8\%$ improvement on WER of dev93 validation set and $12.2\%$ on WER of eval92 testing set. This demonstrated that our models have better generalization ability with training on AXE loss.
The Conformer encoder outperforms the Transformer encoder, and they have equivalent RTF speed. With dynamic rectification method, the Mask + Rec + AXE model could further improve the WER results on both dev93 and eval92 dataset. %The mask probability of rectification $\rho_{rec}=100\%$ has the best performance at our experiments.
\begin{table}[ht]
	\vspace{-1em}
	\centering
	\caption{Different Training Methods and Decoding Iterations, WER and RTF Results on WSJ}
	\scalebox{0.85} {
		\begin{tabular}{p{3.2cm}|p{0.8cm}|p{1.0cm}|p{1.0cm}|p{1.0cm}}
			\hline\hline
			\textbf{Training Method} & \textbf{Iter} & \textbf{dev93} & \textbf{eval92} & \textbf{RTF} \\
			\hline
			\multicolumn{4}{l}{\textbf{\textit{Transformer}}} \\
			\hline
			\quad Mask + CE & 1 & 16.8 & 14.3 & 0.04 \\
			\quad Mask + AXE & 1 & 15.8 & 12.5 & 0.04 \\
			\quad Mask + Rec + AXE & 1 & 15.3 & 11.8 & \textbf{0.04} \\
			\hline
			\quad Mask + CE & 10 & 16.5 & 13.9 & 0.07 \\
			\quad Mask + AXE & 10 & 15.7 & 12.2 & 0.07 \\
			\quad Mask + Rec + AXE & 10 & \textbf{15.2} & \textbf{11.6} & 0.07 \\
			\hline\hline
			\multicolumn{4}{l}{\textbf{\textit{Conformer}}} \\
			\hline
			\quad Mask + CE & 1 & 14.6 & 12.1 & 0.04 \\
			\quad Mask + AXE & 1 & 13.9 & 11.4 & 0.04 \\
			\quad Mask + Rec + AXE & 1 & 13.7 & 11.3 & \textbf{0.04} \\
			\hline
			\quad Mask + CE & 10 & 14.1 & 11.7 & 0.07 \\
			\quad Mask + AXE & 10 & 13.7 & 11.2 & 0.07 \\
			\quad Mask + Rec + AXE & 10 & \textbf{13.6} & \textbf{11.1} & 0.07 \\
			\hline\hline
		\end{tabular}
	}
	\label{tab3}
	\vspace{-1em}
\end{table}
%As listed in Table~\ref{tab4}, we also investigated different parameters in the decoding procedure. We found that the best CTC confidence threshold is $P_{thres}=0.999$ for both Mask + AXE and Mask \& Rectification + AXE model. By dynamic rectification method, the Mask \& Rectification + AXE model needs less masked tokens, and corrects the unmasked high confident tokens at decoding procedure automatically. In addition, with larger decoding iterations, the WER performance of the model will become better. The results also demonstrated that our models are insensitive to the decoding parameters. When the iteration number $K$ is changed from $10$ to $1$, it suffers only a little degradation on WER results. However, it gains a huge of inference speed promotion with relatively $75\%$ improvement on RTF performance.
In addition, we also investigate different parameters in decoding procedure.
%We found that the best CTC confidence threshold is $P_{thres}=0.999$ for all models. By dynamic rectification method, the Mask \& Rectification + AXE model needs fewer masked tokens and corrects the unmasked high confident tokens at decoding procedure automatically.
With more decoding iterations, the WER performance of the model will become better. The results also demonstrated that our models are insensitive to decoding parameters. When the iteration number $K$ is changed from $10$ to $1$, it suffers only a little degradation on WER results. However, it gains a huge inference speed promotion with a relatively $75\%$ improvement on RTF performance.

\subsection{Analysis}
\label{ssec:analysis}
\begin{figure}[!b]
	\centering
	\subfigure[CE Loss] {\label{fig2a}
		\includegraphics[width=0.47\columnwidth]{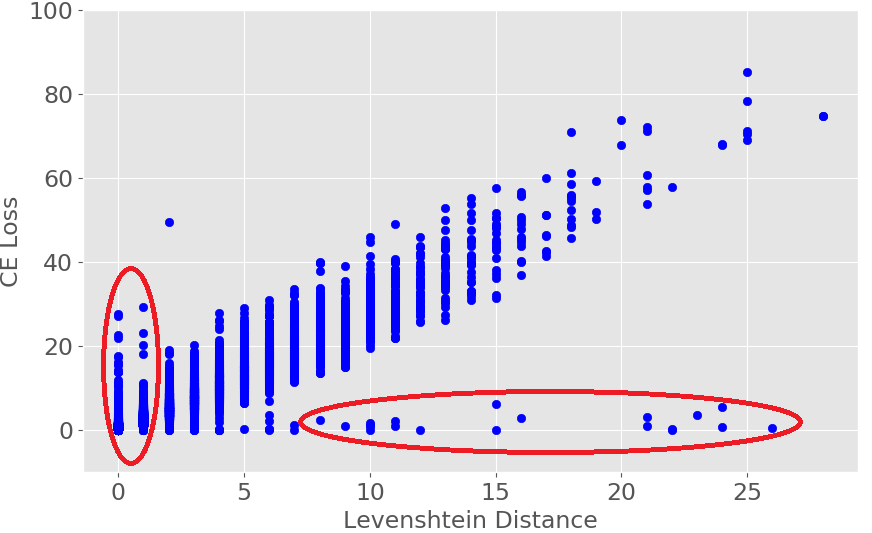}
	}
	\subfigure[AXE Loss] {\label{fig2b}
		\includegraphics[width=0.47\columnwidth]{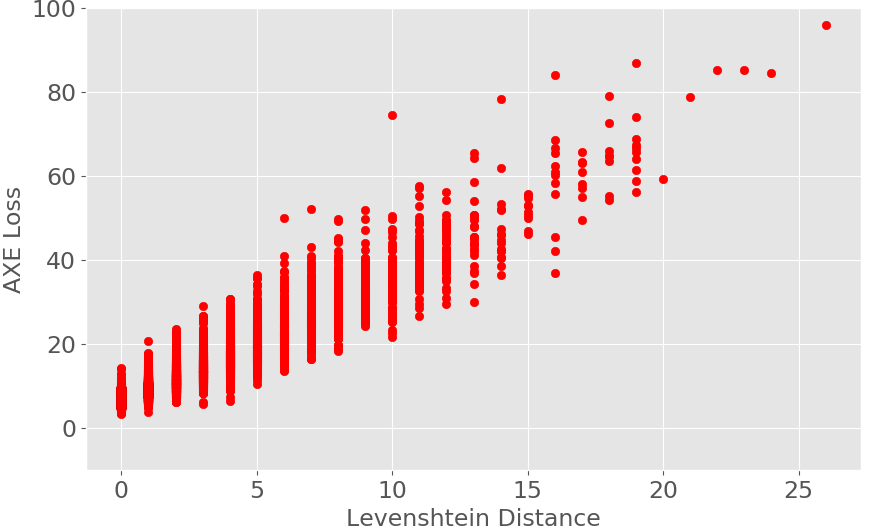}
	}
	\caption{Relationship Between Training Loss and Levenshtein Distance of Ground Truth and Predicted Sentences}
	\label{fig2}
\end{figure}

% We compare our best Mask + Rec + AXE model with other autoregressive and non-autoregressive methods.
We contrast different autoregressive and non-autoregressive approaches with our best Mask + Rec + AXE model.
As shown in Table~\ref{tab4}, our dynamic alignment Mask CTC model achieves better WER results than vanilla Mask CTC* (our results) at both Transformer and Conformer architecture. The results also indicate that our method outperforms most previous non-autoregressive methods.
%For fair comparison, we used the same Transformer model architecture. Therefore, our model has the same RTF results as the Mask CTC and better than DLP Mask CTC.
In addition, DLP Mask CTC needs an additional predictor to estimate the length of masks, resulting in a more complicated model structure but slightly better performance. Our methods could be easily applied to similar length predictions with a lighter structure.

\begin{table}[ht]
	\vspace{-1em}
	\centering
	\caption{Compared with Other Non-Autoregressive and Autoregressive Methods, WER and RTF Results on WSJ}
	\scalebox{0.85} {
		\begin{tabular}{p{4.5cm}|p{0.5cm}|p{0.8cm}|p{0.8cm}|p{0.6cm}}
			\hline\hline
			\textbf{Model} & \textbf{Iter} & \textbf{dev93} & \textbf{eval92} & \textbf{RTF} \\
			\hline
			\multicolumn{5}{l}{\textbf{\textit{Autoregressive}}} \\
			\hline
			%Transformer CTC-Attention ~\cite{Karita2019Improving} & S & 14.4 & 11.3 & 0.97 \\
			\textit{Transformer} & & & & \\
			\quad CTC-Attention & S & 13.5 & 10.9 & 4.62 \\
			\textit{Conformer} & & & & \\
			\quad CTC-Attention & S & 11.1 & 8.5 & 5.09 \\
			\hline
			\hline
			\multicolumn{5}{l}{\textbf{\textit{Non-Autoregressive Previous Work}}} \\
			\hline
			\textit{Transformer} & & & & \\
			\quad CTC & 1 & 19.4 & 15.5 & 0.03 \\
			%Transformer Mask CTC~\cite{higuchi2020mask} & 1 & 15.7 & 12.5 & 0.04 \\
			%\quad Mask CTC & 10 & 14.9 & 12.0 & 0.06 \\
			\quad Mask CTC* & 10 & 16.5 & 13.9 & 0.06 \\
			\quad Mask CTC + DLP & 10 & 13.8 & 10.8 & 0.07 \\
			\quad Imputer (IM) & 8 & - & 16.5 & - \\
			\quad Imputer (DP) & 8 & - & 12.7 & - \\
			\quad Align-Refine & 10 & 13.7 & 11.4 & 0.06 \\
			\textit{Conformer} & & & & \\
			\quad CTC & 1 & 13.0 & 10.8 & 0.03 \\
			%\quad Mask CTC  & 10 & 11.7 & 9.1 & 0.06 \\
			\quad Mask CTC* & 10 & 14.1 & 11.7 & 0.06 \\
			\quad Mask CTC + DLP & 10 & 11.3 & 9.1 & 0.08 \\
			\hline
			\multicolumn{5}{l}{\textbf{\textit{Our Work}}} \\
			\hline
			\textit{Transformer} & & & & \\
			%\textbf{Dynamic Alignment Mask CTC (ours)} & 1 & 15.3 & 11.8 & 0.04 \\
			\quad Proposed & 10 & 15.2 & 11.6 & 0.07 \\
			\textit{Conformer} & & & & \\
			\quad Proposed & 10 & 13.6 & 11.1 & 0.07 \\
			\hline\hline
		\end{tabular}
	}
	\label{tab4}
 \vspace{-1em}
\end{table}

To investigate how AXE loss improves the model training, we plot the scatter diagram between training loss and Levenshtein distance in Figure~\ref{fig2}. The Levenshtein distance is calculated by ground truth sentence and predicted output by iterative decoding with maximum token probability. The figures show that both AXE and CE loss have a linear relationship with Levenshtein distance. However, there are two kinds of outliers for CE loss (see red boxes in Figure~\ref{fig2a}). The first outliers have large CE loss but have small Levenshtein distance, mainly caused by token order mismatch, even if their edit distance is small. By contrary, the AXE will have reasonable loss value by relieving this order mismatch penalty. In addition, the second outliers have small CE loss, but their predictions are quite different from ground truth sentence. We conjecture that multimodality could still be a problem for CE loss. For example, ``form or'' and ``for more'' have similar pronunciations and may have both high token probabilities and low CE losses. Instead, their AXE losses will be quite different (see Figure~\ref{fig2b}). %Therefore, the AXE loss is more related with the model prediction than CE loss, and would lead to training a model with higher accuracy (see Figure~\ref{fig2b}).

% \begin{figure}[ht]
% 	\centering
% 	\subfigure[Without Rectification] {\label{fig3a}
% 		\includegraphics[width=0.47\columnwidth]{figs/fig3a}
% 	}
% 	\subfigure[With Rectification] {\label{fig3b}
% 		\includegraphics[width=0.47\columnwidth]{figs/fig3b}
% 	}
% 	\caption{Compared with Levenshtein Distance Distribution of Training and Inference Decoder Input}
% 	\label{fig3}
% \end{figure}

% To verify the effectiveness of dynamic rectification method, we also compared the levenshtein distance distribution with or without rectification on eval92 set. Without rectification, the edit distance is calculated by greedy CTC output and ground truth sentence (see Figure~\ref{fig3a}). With rectification, the ground truth sentence is replaced with $Y_{rec}$ in Algorithm~\ref{method3} (see Figure~\ref{fig3b}). We found that with rectification the proportion of small distance rise greatly, compared with ones without rectification. It means that rectification method narrows the gap between training and inference decoder input.

\section{Conclusions}
\label{sec:cls}
%In this paper, we proposed an end-to-end NAR speech recognition model, dynamic alignment Mask CTC. The AXE loss makes the model focus on the tokens matching, and relax the restriction for tokens order. The dynamic rectification method could reduce the mismatch of decoder input between training and inference, handling the high confident but possible wrong tokens of greedy CTC output. The experimental results demonstrated that our proposed model achieves not only better WER performance but also faster training convegence, than vanilla Mask CTC model. Future works, including integration of external language model in the non-autoregressive decoding, could be extended. We are also planning to design a length changeable alignment method, breaking the one-to-one restriction of matching decoder input and output tokens, and further reducing the gap of the performance between NAR and AR models in ASR tasks.
In this paper, we propose an end-to-end NAR speech recognition model, dynamic alignment Mask CTC. The AXE loss makes the model focus on the tokens matching and relaxes the restriction of tokens order. The dynamic rectification could reduce the mismatch of decoder input between training and inference, simulating the high confidence but possible wrong tokens of greedy CTC output. 
% The experimental results demonstrate that our proposed model achieves better WER performance than the vanilla Mask CTC model. 
Experimental results demonstrate that our proposed model outperforms the vanilla Mask CTC model in terms of WER.
%Future works, including integration of external language model in the non-autoregressive decoding, could be extended.
% In future works, we plan to design a length changeable alignment method, breaking the one-to-one restriction of matching decoder input and output tokens. Combining these methods could further reduce the performance gap between NAR and AR models in ASR tasks.
\section{Acknowledgement}
Supported by the Key Research and Development Program of Guangdong Province (grant No. 2021B0101400003) and Corresponding author is Jianzong Wang (jzwang@188.com).

\bibliographystyle{IEEEbib}
\bibliography{MaskCTC}

\begin{thebibliography}{10}

\bibitem{Song2021Non}
Xingchen Song, Zhiyong Wu, Yiheng Huang, Chao Weng, Dan Su, and Helen Meng,
\newblock ``Non-autoregressive transformer asr with ctc-enhanced decoder
  input,''
\newblock in {\em ICASSP}, 2021, pp. 5894--5898.

\bibitem{Fan2021CASS}
Ruchao Fan, Wei Chu, Peng Chang, and Jing Xiao,
\newblock ``Cass-nat: Ctc alignment-based single step non-autoregressive
  transformer for speech recognition,''
\newblock in {\em ICASSP}, 2021, pp. 5889--5893.

\bibitem{Tian2020Spike}
Zhengkun Tian, Jiangyan Yi, Jianhua Tao, Ye~Bai, Shuai Zhang, and Zhengqi Wen,
\newblock ``Spike-triggered non-autoregressive transformer for end-to-end
  speech recognition,''
\newblock in {\em INTERSPEECH}, 2020, pp. 5026--5030.

\bibitem{Bai2020Listen}
Ye~Bai, Jiangyan Yi, Jianhua Tao, Zhengkun Tian, Zhengqi Wen, and Shuai Zhang,
\newblock ``Listen attentively, and spell once: Whole sentence generation via a
  non-autoregressive architecture for low-latency speech recognition,''
\newblock in {\em INTERSPEECH}, 2020, pp. 3381--3385.

\bibitem{Majumdar2021Citrinet}
Somshubra Majumdar, Jagadeesh Balam, Oleksii Hrinchuk, Vitaly Lavrukhin, Vahid
  Noroozi, and Boris Ginsburg,
\newblock ``Citrinet: Closing the gap between non-autoregressive and
  autoregressive end-to-end models for automatic speech recognition,''
\newblock in {\em arXiv preprint arXiv:2104.01721}, 2021.

\bibitem{graves2006connectionist}
Alex Graves, Santiago Fern{\'a}ndez, Faustino Gomez, and J{\"u}rgen
  Schmidhuber,
\newblock ``Connectionist temporal classification: labelling unsegmented
  sequence data with recurrent neural networks,''
\newblock in {\em ICML}, 2006, pp. 369--376.

\bibitem{zhang2021wenet}
Binbin Zhang, Di~Wu, Chao Yang, Xiaoyu Chen, Zhendong Peng, Xiangming Wang,
  Zhuoyuan Yao, Xiong Wang, Fan Yu, Lei Xie, et~al.,
\newblock ``Wenet: Production first and production ready end-to-end speech
  recognition toolkit,''
\newblock in {\em INTERSPEECH}, 2021, pp. 4054--4058.

\bibitem{Devlin2019BERT}
Ming-Wei Devlin, Jacoband~Chang, Kenton Lee, and Kristina Toutanova,
\newblock ``Bert: Pre-training of deep bidirectional transformers for language
  understanding,''
\newblock in {\em IEEE International Conference of the North American Chapter
  of the Association for Computational Linguistics (NAACL)}, 2019, pp.
  4171--4186.

\bibitem{ghazvininejad2019mask}
Marjan Ghazvininejad, Omer Levy, Yinhan Liu, and Luke Zettlemoyer,
\newblock ``Mask-predict: Parallel decoding of conditional masked language
  models,''
\newblock in {\em EMNLP}, 2019, pp. 6111--6120.

\bibitem{ghazvininejad2020semi}
Marjan Ghazvininejad, Omer Levy, and Luke Zettlemoyer,
\newblock ``Semi-autoregressive training improves mask-predict decoding,''
\newblock in {\em arXiv preprint arXiv:2001.08785}, 2020.

\bibitem{chan2020imputer}
William Chan, Chitwan Saharia, Geoffrey Hinton, Mohammad Norouzi, and Navdeep
  Jaitly,
\newblock ``Imputer: Sequence modelling via imputation and dynamic
  programming,''
\newblock in {\em ICML}, 2020, pp. 1403--1413.

\bibitem{higuchi2020mask}
Yosuke Higuchi, Shinji Watanabe, Nanxin Chen, Tetsuji Ogawa, and Tetsunori
  Kobayashi,
\newblock ``Mask ctc: Non-autoregressive end-to-end asr with ctc and mask
  predict,''
\newblock in {\em INTERSPEECH}, 2020, pp. 6112--6121.

\bibitem{higuchi2020improved}
Yosuke Higuchi, Hirofumi Inaguma, Shinji Watanabe, Tetsuji Ogawa, and Tetsunori
  Kobayashi,
\newblock ``Improved mask-ctc for non-autoregressive end-to-end asr,''
\newblock in {\em ICASSP}, 2020, pp. 8363--8367.

\bibitem{Gulati2020Conformer}
Anmol Gulati, James Qin, Chung-Cheng Chiu, Niki Parmar, Yu~Zhang, Wei Han,
  Shibo Wang, Zhengdong Zhang, Yonghui Wu, and Ruoming Pang,
\newblock ``Conformer: Convolution-augmented transformer for speech
  recognition,''
\newblock in {\em INTERSPEECH}, 2020, pp. 5036--5040.

\bibitem{Ghazvininejad2020Aligned}
Marjan Ghazvininejad, Vladimir Karpukhin, Luke Zettlemoyer, and Omer Levy,
\newblock ``Aligned cross entropy for non-autoregressive machine translation,''
\newblock in {\em ICML}, 2020, pp. 3515--3523.

\bibitem{vaswani2017attention}
Ashish Vaswani, Noam Shazeer, Niki Parmar, Jakob Uszkoreit, Llion Jones,
  Aidan~N. Gomez, Lukasz Kaiser, and Illia Polosukhin,
\newblock ``Attention is all you need,''
\newblock in {\em NIPS}, 2017.

\bibitem{linguistic1994wsj1}
Linguistic~Data Consortium,
\newblock ``Csr-ii (wsj1) complete,''
\newblock in {\em Linguistic Data Consortium, Philadelphia, vol. LDC94S13A},
  1994.

\bibitem{linguistic2007wsj0}
John Garofalo, David Graff, Doug Paul, and David Pallett,
\newblock ``Csr-i (wsj0) complete,''
\newblock in {\em Linguistic Data Consortium, Philadelphia, vol. LDC93S6A},
  2007.

\bibitem{Watanabe2021ESPnet}
Shinji Watanabe, Florian Boyer, Xuankai Chang, Pengcheng Guo, Tomoki Hayashi,
  Yosuke Higuchi, Takaaki Hori, Wen-Chin Huang, Hirofumi Inaguma, Naoyuki Kamo,
  Shigeki Karita, Chenda Li, Jing Shi, Aswin Subramanian, and Wangyou Zhang,
\newblock ``The 2020 espnet update: New features, broadened applications,
  performance improvements, and future plans,''
\newblock in {\em IEEE Data Science and Learning Workshop (DSLW)}, 2021.

\bibitem{Tom2015Audio}
Tom Ko, Vijayaditya Peddinti, Daniel Povey, and Sanjeev Khudanpur,
\newblock ``Audio augmentation for speech recognition,''
\newblock in {\em INTERSPEECH}, 2015.

\bibitem{Daniel2019SpecAugment}
Daniel~S. Park, William Chan, Yu~Zhang, Chung-Cheng Chiu, Barret Zoph, Ekin~D.
  Cubuk, and Quoc~V. Le,
\newblock ``Specaugment: A simple data augmentation method for automatic speech
  recognition,''
\newblock in {\em INTERSPEECH}, 2019, pp. 2613--2617.

\end{thebibliography}

\end{document}